\documentclass[showpacs,preprintnumbers,prd,nofootinbib,floats,amssymb,floatfix]{revtex4}
\usepackage{amsmath}
\usepackage{amsfonts}
\usepackage{graphicx}
\usepackage{hyperref}

\usepackage{amsmath}
\usepackage{amstext}
 
\begin{document}
\title{Canonical and Symplectic  analysis of actions  describing linearized gravity }
\author{ Alberto Escalante} \email{aescalan@ifuap.buap.mx}
 \affiliation{  Instituto de F{\'i}sica Luis Rivera Terrazas, Benem\'erita Universidad Aut\'onoma de Puebla,  \\
 Apartado Postal J-48 72570, Puebla Pue., M\'exico. }
\author{Melissa Rodr{\'i}guez-Z\'arate}  
\affiliation{ Facultad de Ciencias F\'{\i}sico-Matem\'{a}ticas, Benem\'erita Universidad Au\-t\'o\-no\-ma de Puebla,
 Apartado postal 1152, 72001 Puebla, Pue., M\'exico.}

\begin{abstract}
By using the canonical and symplectic approaches  an (nonstandard) alternative action  describing linearized gravity is studied.  We identify the complete set of Dirac's constraints,  the  counting of physical degrees of freedom is performed and the Dirac brackets are constructed. Furthermore, the symplectic analysis is developed which includes  the complete set of Faddeev-Jackiw constraints and  a symplectic tensor;  from that symplectic   matrix we show that the generalized Faddeev-Jackiw brackets and  the Dirac ones  coincide to each other. With all these results at hand, we prove  that   the number of physical degrees of freedom are eight, thus,    we conclude that the   theory does not describe the dynamics of  linearized gravity. In addition, we also develop the symplectic analysis of standard linearized gravity and we compare the results for both standard and nonstandard theories.   Finally we present some  remarks and conclusions.   
\end{abstract}
\date{\today}

 \maketitle
\preprint{}
\section{INTRODUCTION}
It is well-known that the study of singular theories,   as  for instance gauge theories,   is a fundamental work to perform. In fact, a gauge theory   within  the Dirac terminology  is  characterized for the presence of     first class constraints which  are  the generators of gauge transformations \cite{1a}.   Furthermore, a gauge  symmetry  is   present in all  fundamental interactions of  nature,  and the study  of that symmetry by using  the  tools developed by Dirac has    allowed us to understand the behavior   of the fundamental interactions within either the  classical or quantum context   \cite{2a}.  In this respect,  gravity can be understood as a gauge theory with the diffeomorphisms  as its gauge symmetry; hence  the extended Hamiltonian of the theory is a linear combination of first class constraints  and performing the counting of physical degrees of freedom,  we obtain  two physical degrees of freedom \cite{3a, 4a}. However, the study of gravity in either the classical or quantum context is a difficult task to develop because   the nonlinearly of the gravitational field is manifested in the  constraints, so far    the  quantization of the theory has not been performed in a  full and satisfactory way \cite{5a, 6a, 7a, 8a}.  In  this respect, in order to obtain new insights in classical or quantum  analysis  of gravity, it is common to study    Einstein's  modified theories    with the goal to provide  new ideas for understanding the symmetries of the gravitational field, examples of such  theories are the so-called     linearized gravity and massive gravity theories. In fact, these theories are   good laboratories  for testing classical and quantum ideas of gravity, for instance, with the detection of gravitational waves made   by  LIGO$\setminus$VIRGO laser interferometers \cite{virgo}, all our actual vision of the universe can change because that discovery could give us a new insight to explore the warper side of the Universe, and it could be an  important starting  point  for testing  new classical or  quantum gravity proposals by means of new actions and its  symmetries. In this respect,   it is well-known that linearized gravity  is a gauge theory that  describes the propagation of a helicity-2 particle  (massless graviton);  from  the Hamiltonian point of view, the theory presents only a set of  first class constraints   (then it is possible fixing the gauge)  and the number of  physical degrees  of freedom is  two in four dimensions. Nevertheless, within  the quantum context,   it is well-known that the usual scheme  of quantization applied to the theory  lead  to infinities that cannot be eliminated by means of regularization and renormalization procedures \cite{barcelos}. Furthermore, one of the alternatives to possibly explain unresolved problems in cosmology  such as the problem of  acceleration of the Universe is performing a  modification of linearized gravity. In this regard, the most natural  modification is promoting  the  helicity-two  theory to  one of a massive spin-two, this theory is the so-called massive gravity \cite{mass1}.  In this manner, any alternative or modified  theory of gravity is an important achievement  and it is mandatory to investigate  the new proposal and their  symmetries  in a  full detail. \\
With these motivations, in this paper we study an alternative theory for linearized gravity that  was proposed in \cite{r1, r2};  the model consists  in an action whose   dynamical variables are not the perturbation of the metric  but   an  electric and magnetic-like  fields. Moreover,  the analysis reported in those works was performed ignoring that the action corresponds to a singular system, therefore the principal symmetries of the theory were not reported. In  order to analyze  a singular theory, there are two approaches to obtain in a systematic way the symmetries and observables of any  physical theory: Dirac's formalism \cite{r1} and the Faddeev-Jackiw [FJ] method \cite{ F1}. The former allows  us  to know the complete set of constraints of the theory, namely, first class constraints and second class ones. As a consequence, the physical degrees of freedom can be exactly counted and the relevant symmetries can be obtained. In fact, first class constraints are the generators  of gauge symmetry,  and the second class constraints are  used for constructing the fundamental Dirac's brackets,  which are useful    for identifying  observables. On the other hand,  the FJ method provides a symplectic description  for singular systems,   where   the  basic feature of this approach is the construction of a symplectic tensor;  from the symplectic tensor it is possible to obtain relevant physical information such as, the degrees of freedom, the gauge symmetry  and the quantization brackets (the so-called generalized FJ brackets) can also be obtained. Moreover, in this framework is not necessary to classify the constraints into first  and second class ones; this fact makes the  FJ method more economical than the Dirac scheme. Hence, in this paper  by using the  Dirac and FJ approaches  the action reported in \cite{r1, r2}  is analyzed.  We report in our analysis the complete set of constraints  and the Dirac brackets are constructed, also we report that the proposed theory  is neither  Lorentz nor gauge invariant, thus, these results modify the number of degrees of freedom,  which we obtain eight  physical degrees. In addition, by using the FJ method we report the complete set of FJ constraints, then the  generalized FJ brackets are constructed  and the counting of physical degrees of freedom is performed; the generalized brackets and the Dirac ones coincide to each other.  Finally, we have added at the end of the paper the symplectic analysis of standard linearized gravity which is absent in the literature. We compare the results of  both standard and nonstandard theories and we comment  their differences.   \\
The paper is organized as follows. In the Sec.I  the Hamiltonian analysis  is performed. We identify the full set of constraints which are  all of second class, there are not first class constraints,  therefore we conclude that the action under study is not a gauge theory.  Moreover, by eliminating the second class constraints we calculate the fundamental Dirac's brackets  and  the counting of physical degrees of freedom is carried out. In Sec.II the symplectic analysis  is developed, we obtain the full set of FJ constraints, then  a symplectic tensor is constructed and  the fundamental  FJ brackets are identified. In addition, in the Sec. III the symplectic analysis of standard linearized gravity is performed and we compare  the results of this section with those obtained previously.   Finally,  we present  a summary and conclusions. 
\section{Hamiltonian analysis}
It is well-known that the cornerstone of linearized gravity  is based in considering a perturbation of the fundamental metric  around the Minkowski spacetime, say 
\begin{equation}
g_{\alpha \beta}=\eta_{\alpha \beta} + h_{\alpha \beta}, 
\label{1}
\end{equation}
where the $h_{\alpha \beta}$ represents a small deviation of the fundamental  metric and the background space-time  is Minkowskian with metric $\eta_{\alpha \beta} = (-1, 1, 1, 1)$, here Greek indices run from 0 to 3. By introducing   the metric (\ref{1})   into  the Einstein-Hilbert action  given by 
\begin{equation}
S[g_{\mu \nu}] = \int \sqrt{-g} R dx^4, 
\end{equation}
where $g$  is the metric and $R$ the Ricci scalar,  and just keeping free fields for  $h_{\alpha \beta}$, then   the Lagrangian for   standard linearized gravity   is obtained     \cite{barcelos}
\begin{equation}
L=\frac{1}{4} \partial_\lambda h_{\mu \nu} \partial^{\lambda} h^{\mu \nu} - \frac{1}{4}\partial_\lambda h^{\mu}{_{\mu}} \partial^\lambda  h^{\nu}{_{\nu}} +\frac{1}{2} \partial_\lambda h^{\lambda}_{{\mu}} \partial^{\mu} h^{\nu}{_{\nu}} - \frac{1}{2} \partial_{\lambda}h^{\lambda} _{\mu} \partial_{\nu} h^{\nu \mu}.
\label{Llin}
\end{equation}
The Lagrangian given in (\ref{Llin}) has been studied in the literature (see \cite{barcelos} and cites there in). In fact, the theory   describes the propagation of a massless particle (the graviton) with two physical degrees of freedom in four dimensions, and from the Hamiltonian point of view, the theory is a gauge theory; there are only first class constraints, then in order to calculate the Dirac brackets   the gauge  is fixed  and the constraints are now converted into  second class constraints. In this respect,  we will  see along this paper that   these symmetries  will not be present in the theory proposed in \cite{r1, r2}. \\
Hence, a new  Lagrangian for  linearized gravity was proposed in  \cite{r1, r2}.  In fact, by considering the tensor field 
\begin{equation}
K_{\alpha \beta \gamma \delta}= \frac{1}{2} \left[ \partial_\alpha  \partial_\gamma  h_{\beta \delta} - \partial_\beta   \partial_\gamma  h_{\alpha\delta} +\partial_\beta  \partial_\delta  h_{\alpha \gamma}  - \partial_\alpha \partial_\delta  h_{\beta \gamma}      \right], 
\end{equation}
which   satisfies 
\begin{eqnarray}
K_{\alpha \beta \gamma \delta}&= &-K_{ \beta \alpha \gamma \delta}= - K_{\alpha \beta  \delta \gamma} = K_{  \gamma \delta \alpha \beta}, \nonumber \\
K_{\alpha \beta \gamma \delta} &+& K_{\alpha  \delta \beta \gamma} + K_{\alpha  \gamma \delta \beta }=0,\nonumber \\
\partial_\alpha K_{ \beta \gamma \delta \epsilon} &+&  \partial_\epsilon K_{ \beta \gamma \alpha \delta } +  \partial_\delta  K_{ \beta \gamma \epsilon \alpha  }=0, \label{eq5}
\end{eqnarray}
where the second and third relations can be identified with the Ricci and Bianchi identities respectively. Moreover,  it can be showed that the linearized Einstein vacuum  field equations are given by  (see \cite{r1, r2, r3} for full details)
\begin{equation}
K_{\alpha \beta }=0, 
\end{equation}
and this implies that all the components $K_{\alpha \beta \gamma \delta}$ can be expressed in terms of the fields $E_{ij}$  and $B_{ij}$ defined by 
\begin{equation}
E_{ij}\equiv K_{0i0j}, \quad \quad B_{ij}\equiv - K^{*}_{0i0j}, 
\end{equation}
where $K^{*} _{\alpha \beta \gamma \delta} = \frac{1}{2} K{_{\alpha \beta }}^{\rho \sigma} \epsilon_{\rho \sigma \gamma \delta}$ is  the dual of $K_{\alpha \beta \gamma \delta}$ and the fields $E_{ij}$, $B_{ij}$ have vanishing  trace.  Then, in \cite{r1, r2} the following action  for describing the linearized  Einstein vacuum   equations  is proposed 
\begin{eqnarray}
S[E, B]&=& \int  \left[ B^{ij}\left(\frac{1}{c}\partial_{t}E_{ij} - \epsilon^{kl}{_i}\partial_{k}B_{lj}\right) - E^{ij}\left(\frac{1}{c}\partial_{t}B_{ij} + \epsilon^{kl}{_i}\partial_{k}E_{lj}\right)\right] dx^4,
\label{ac1}
\end{eqnarray} 
where $i,j,k= 1,2,3$.  The equations of motion obtained from the action (\ref{ac1}) are given by 
\begin{equation}
\frac{1}{c} \partial_t E_{ij}= \epsilon_{i}{^{kl}} \partial_k B_{lj}, \quad \quad \frac{1}{c} \partial_t B_{ij}= - \epsilon_{i}{^{kl}} \partial_k E_{lj}, 
\label{mov1}
\end{equation}
and the fields $E_{ij}$ and $B_{ij}$ satisfy  the following constraints  
\begin{align}
\partial_{i}E^{ij}= 0,	&&
\partial_{i}B^{ij}= 0, 
\label{mov2}
\end{align}
equations (\ref{mov1}) and (\ref{mov2}) are the equations of motion of linearized gravity given in non-covariant way. As it was commented above, the  alternative action (\ref{ac1}) was studied in \cite{r1, r2, r3}, however we shall show   that the  action  is a singular system and this fact was ignored in those works. It is important to note that  the action (\ref{ac1})  reproduces   only the equations (\ref{mov1});   the equations  (\ref{mov2})  are obtained from the equations (\ref{eq5}) and they were added   by hand  in \cite{r1, r2, r3}   in order to obtain a set of equations similar  to Maxwell's theory, however, we will observe  that these facts yield different symmetries  of those known  for  linearized gravity.  In this manner,   in order to perform a complete study of the action  (\ref{ac1})  we will use the Dirac formulation for  constrained systems. It is important to comment that the action (\ref{ac1})  is neither Lorentz invariant  nor   gauge theory; these facts  will be reflected in the Hamiltonian analysis, in particular in the number of physical degrees of freedom of the theory. \\
We can observe  that the matrix elements of the Hessian given by
\begin{align*} 
\frac{{\partial^{2}{\mathcal{L}}}}{{\partial(\partial_{t} E_{ij})\partial(\partial_{t} E_{kl})}}, &&
\frac{{\partial^{2}{\mathcal{L}}}}{{\partial(\partial_{t} E_{ij})\partial(\partial_{t} B_{kl})}}, &&
\frac{{\partial^{2}{\mathcal{L}}}}{{\partial(\partial_{t} B_{ij})\partial(\partial_{t} B_{kl})}},
\end{align*}
are identically zero, hence the system is singular and  we expect  primary constraints. In order to identify the primary constraints, the canonical formalism calls for the definition of the momenta $(P^{ij}, \Pi^{ij})$ canonically conjugate to $(E_{ij}, B_{ij})$ are given by
\begin{align*}
P^{ij} = \frac{\delta{\mathcal{L}}}{\delta \dot{E}_{ij}}, &&
\Pi^{ij} = \frac{\delta{\mathcal{L}}}{\delta \dot{B}_{ij}}.
\end{align*} 
In this manner, the fundamental Poisson brackets are 
\begin{eqnarray}
\{ E_{ij} (x), P^{kl} (y) \} = \frac{1}{2} \left( \delta^k_i \delta^l_j + \delta^k_j \delta^l_i  \right) \delta^3(x-y), \nonumber \\
\{ B_{ij} (x), \Pi^{kl} (y) \} = \frac{1}{2} \left( \delta^k_i \delta^l_j + \delta^k_j \delta^l_i  \right) \delta^3(x-y).
\end{eqnarray}
From  the definition of the momenta, we identify the following 10 primary constraints
\begin{align}
\chi^{ij} : P^{ij} - \frac{1}{c}B^{ij} \approx 0, &&
\bar{\chi}^{ij} :\Pi^{ij} + \frac{1}{c}E^{ij} \approx 0,
\label{cons}
\end{align} 
these primary constraints are of second class and their evolution in time will fix the Lagrange multipliers; for this theory there are not more constraints. It is important to note that there are not first class constraints and therefore the theory under study is not a gauge theory. In fact, this result does not  agree with the gauge invariance that is present in  the standard   Lagrangian of linearized gravity. \\
On the other hand, because  of there are second class constraints,  we will construct the Dirac brackets from  the following matrix whose entries are given by the Poisson brackets between the second class constraints 
\begin{eqnarray*}
\label{eq}
C_{\alpha \beta} =
\left(
  \begin{array}{cccc}
   0 	& \quad 	-\frac{1}{c}\left(\eta^{ik}\eta^{jl} + \eta^{il}\eta^{jk}\right)                 \\
   \frac{1}{c}\left(\eta^{ik}\eta^{jl} + \eta^{il}\eta^{jk}\right)	  &\quad    	0 \\
      \end{array}
\right)\delta^{3}(x-y),
\end{eqnarray*}
\\
its inverse is given by
\begin{eqnarray*}
\label{eq}
C^{-1}_{\alpha \beta} =
\left(
  \begin{array}{cccc}
  0	  &\quad  	\frac{c}{4}\left(\eta_{ik}\eta_{jl} + \eta_{il}\eta_{jk}\right)                                                                           \\
    -\frac{c}{4}\left(\eta_{ik}\eta_{jl} + \eta_{il}\eta_{jk}\right)	  &\quad  	  0 \\
      \end{array}
\right)\delta^{3}(x-y).
\end{eqnarray*} \\

Furthermore, the Dirac brackets between two functionals, say $A,B$ are expressed by
$\{A(x),B(y)\}_{D} = \{A(x),B(y)\} - \int dudv\{A(x),\chi_{\alpha}(u)\}C^{-1}_{\alpha\beta}\{\chi_{\beta}(v),B(y)\},$ where $\{A(x),B(y)\}$ is the usual Poisson bracket between the functionals $(A,B)$ and $(\chi_{\alpha},\chi_{\beta})$ represent the set of second class constraints. Hence, we obtain the following Dirac's brackets of the theory

\begin{eqnarray*}
\{E_{ij}(x), P^{kl}(y)\}_{D} &=& \frac{1}{4}\left(\delta^{k}_{i}\delta^{l}_{j} + \delta^{k}_{j}\delta^{l}_{i}\right)\delta^{3}(x-y), 	\\
\{B_{ij}(x), \Pi^{kl}(y)\}_{D} &=& \frac{1}{4}\left(\delta^{k}_{i}\delta^{l}_{j} + \delta^{k}_{j}\delta^{l}_{i}\right)\delta^{3}(x-y). 
\end{eqnarray*}
With  these results at hand, we are able to calculate the physical degrees
of freedom as follows; there are 20 canonical  variables and 10 second class constraints, thus, there are five  physical degrees of freedom. Nonetheless, we need
to take into the account the equations (\ref{mov2}). Hence,  the constraints (\ref{cons}) satisfy the following  reducibility conditions
\begin{eqnarray}
\partial_i  \chi^{ij}&=&0, \nonumber\\
\partial_i \bar{\chi}^{ij}&=&0, 
\end{eqnarray}
which  imply that there are [10-6]=4 second class constraints, therefore,  the physical degrees of freedom are eight.  It is important to remark that our results indicate that in
spite of the action (\ref{ac1}) yields linearized Einstein's  equations of motion,  it does not describe the dynamics of linearized gravity  at all. In fact, it is well-known that standard linearized gravity  is  both  gauge invariant and  describes the propagation of a massless particle  with  two degrees of freedom. In this manner, we have found strong differences  between the action (\ref{ac1}) and  the standard  theory for describing linearized gravity.  \\
We finish this section with some extra comments. The Hamiltonian of the theory is given by 
\begin{equation}
H= \int \left[c \epsilon^{kl}{_i} B^{ij}\partial_{k}P_{lj} - c\epsilon^{kl}{_i}E^{ij}\partial_{k}\Pi_{lj}  \right] dx^3.
\label{ham}
\end{equation}
It is straightforward  to prove    that the Hamiltonian is of first class. In fact, we have 
\begin{eqnarray}
\{ H, \chi^{qr} \}&=& \frac{1}{2}c \epsilon^{kq}{_i} \partial_k \bar{\chi}^{ir}+\frac{1}{2} c \epsilon^{kr}{_i} \partial_k \bar{\chi}^{iq}\approx0, \nonumber \\
\{ H, \bar{\chi}^{qr} \}&=&\frac{1}{2} c \epsilon^{kiq} \partial_k {\chi}{_i}^r +\frac{1}{2} c \epsilon^{kir} \partial_k \chi{_i}^q \approx 0.  
\end{eqnarray}
In this manner, the Hamiltonian found in \cite{r1, r2, r3} is not equivalent to that found in   (\ref{ham}). In these works  was ignored that the theory under study is singular, thus, we can not talk neither first class constraints nor second class constraints. However, we have showed that the Hamiltonian (\ref{ham}) is of first class and therefore within  Dirac's terminology it is an observable. All results found in this section extend and complete the work  reported in \cite{r1, r2, r3}. 
 %%%%%%%%%%%%%%%%%%%%%%%%%%%%%%%%%%%%%%%%%%
\section{Symplectic Analysis.}
Now we will reproduce the  results obtained within the Dirac scheme  by using  the FJ  analysis. We start with the Lagrangian (\ref{ac1}) rewritten as 
\begin{eqnarray}
{\mathcal{L}} &=& \frac{2}{c}B^{ij}\dot{E}_{ij} - \frac{2}{c}E^{ij}\dot{B}_{ij} - 2\epsilon^{kl}{_i} B^{ij}\partial_{k}B_{lj} -2\epsilon^{kl}{_i}E^{ij}\partial_{k}E_{lj}, 
\label{sim1}
\end{eqnarray}
note that we  have included an additional factor of 2. This overall factor will not affect the Euler-Lagrange equations of motion in any way. Hence, from the Lagrangian (\ref{sim1}) we obtain the following symplectic Lagrangian 
\begin{equation}
\overset{(0)}{\mathcal{L}} = 2P^{ij}\dot{E}_{ij} + 2\Pi^{ij}\dot{B}_{ij} - \overset{(0)}{V},
\end{equation}
where $\overset{(0)}{V} = 2c \epsilon^{kl}{_i} B^{ij}\partial_{k}P_{lj} - 2c\epsilon^{kl}{_i}E^{ij}\partial_{k}\Pi_{lj}$ is identified as  the symplectic potential. Furthermore, in the FJ framework, the Euler-Lagrange equations of motion  are given by  \cite{F2}
\begin{equation}
f^{(0)}_{ab}\dot{\xi}^{b}=\frac{\partial V^{(0)}(\xi)}{\partial\xi^{a}},
\label{eq36}
\end{equation}
where the symplectic matrix $f^{(0)}_{ab}$ takes the form
\begin{equation}
f^{(0)}_{ab}(x,y)=\frac{\delta \mathrm{a}_{b}(y)}{\delta\xi^{a}(x)}-\frac{\delta \mathrm{a}_{a}(x)}{\delta\xi^{b}(y)},
\label{37}
\end{equation}
with $\xi{^{(0)}}=\left(E_{ij}, P^{ij}, B_{ij}, \Pi^{ij} \right)$ and $\mathrm{a}{^{(0)}}= \left(2P^{ij}, 0, 2\Pi^{ij},0 \right)$ representing  a set of symplectic variables. The matrix (\ref{37}) is not singular and this implies that there are not FJ constraints. However, as it was commented above we need to take into account the conditions (\ref{mov2}),  which   implies that $\partial_{i}P^{ij}=0$, $\partial_{i}\Pi^{ij}=0$ and  they will be considered as constraints.  Moreover, we can see that $\partial_{j}\partial_{i}P^{ij}=0$, and $\partial_{j}\partial_{i}\Pi^{ij}=0$ which correspond to. All this  information must  be added to the symplectic Lagrangian by using Lagrange multipliers,  namely,  $\gamma_i$ and $\sigma_i$, thus we obtain
 \begin{equation}
 \overset{(1)}{\mathcal{L}} = 2P^{ij}\dot{E}_{ij} + 2\Pi^{ij}\dot{B}_{ij} - \left[2\partial_{i}P^{ij}-\rho^{j}\right]\dot{\gamma}_{j} - \left[2\partial_{i}\Pi^{ij} - \alpha^{j}\right]\dot{\sigma}_{j} - \overset{(0)}{V}.
 \label{sim2}
 \end{equation}
Because of the reducibility conditions,   we have added the Lagrange multiplier of the Lagrange multiplier, $\rho^i$ and $\alpha^i$,  and with this fact  under consideration, we can obtain a symplectic tensor \cite{F2}. From (\ref{sim2}) we can identify the following symplectic variables $\overset{(1)}{\xi} =\left(E_{ij}, P^{ij}, B_{ij}, \Pi^{ij}, \gamma_{j}, \sigma_{j}, \rho^{j}, \alpha^{j}\right)$ and the following  1-forms
$\overset{(1)}{a} =\left(2P^{ij}, 0, 2\Pi^{ij}, 0, -\left[2\partial_{i}P^{ij} - \rho^{j}\right], - \left[2\partial_{i}\Pi^{ij} - \alpha^{j}\right], 0, 0\right)$. In this manner, by using the symplectic variables and the 1-forms,  we obtain the following symplectic matrix 
\begin{eqnarray*}
\label{eq}
\overset{(1)}{f}_{ij}=
\left(
  \begin{array}{cccccccc}
0	  &\,\,	-\delta^{ij}_{kl}	  &\,\,	 	 0 	&\,\,		 0	&\,\,		 0 	&\,\,	 0         &\,\,		 0            &\,\,		 0                                                                       
  \\                                                           
\delta^{ij}_{kl}	  &\,		0  	&\, 	0  	&\,  		0	&\,	-\left(\delta^{i}_{l}\partial_{k} + \delta^{i}_{k}\partial_{l}\right)		&\,		 0	&\, 		 0  	&\,		 0  
\\                                                               
 0	&\, 		 0 	&\, 		 0 	&\, 	-\delta^{ij}_{kl}	  	&\,	0	&\,	 0 	&\, 		 0 	&\, 		 0 		
\\
0	&\, 		 0 	&\, 		 \delta^{ij}_{kl}	&\,		0	&\, 	0	&\,	-\left(\delta^{i}_{l}\partial_{k} + \delta^{i}_{k}\partial_{l}\right)		&\, 		0	&\, 	0
\\
0	&\, 	\left(\delta^{i}_{l}\partial_{k} + \delta^{i}_{k}\partial_{l}\right)	&\, 	0	&\, 	0	&\, 	0	&\, 	0	&\, 	-\delta^{i}_{j}		&\, 		0
\\
0	&\, 	0	&\,	0	&\, 	\left(\delta^{i}_{l}\partial_{k} + \delta^{i}_{k}\partial_{l}\right)	&\, 	0	&\,	0	&\, 	0	&\,	-\delta^{i}_{j}
\\
0	&\, 	0	&\, 	0	&\,	0	&\,	\delta^{i}_{j}		&\, 	0	&\,	0	&\,	0
\\
0	&\,	0	&\, 	0	&\, 	0	&\,	0	&\,	\delta^{i}_{j}		&\,	0	&\, 	0
      \end{array}
\right)\delta^3(x-y),
\end{eqnarray*}
where we can observe that is not a singular matrix, therefore it is a symplectic tensor. The  inverse of $\overset{(1)}{f}_{ij}$ is given by
\begin{eqnarray*}
\label{eq}
\overset{(1)}{f}^{-1}_{ij}&=&
\left(
  \begin{array}{cccccccc}
0	  &		\frac{1}{4}(\delta^{i}_{k}\delta^{j}_{l} +  \delta^{i}_{l}\delta^{j}_{k}	)		  &\quad\quad		 	 0 	&\quad\quad			 0	&\quad\quad		 0 	&\quad\quad		 0         &\quad\quad			                                                                             
\\            
 	&\quad			&\quad		&\quad       &\quad         &\quad	&\quad		&\quad	                                                                  
\\                             
-\frac{1}{4}(\delta^{i}_{k}\delta^{j}_{l} +\delta^{j}_{k}\delta^{i}_{l})	  &\quad\quad			0  	&\quad\quad	 	0  	&\quad\quad	 		0	&\quad\quad		0	&\quad\quad		 0	&\quad\quad	 				   
\\                                                               
	& \,	&\,	&\,        &\,          &\,	&\,	&\,
\\
 0	&\quad\quad	 		 0 	&\quad\quad	 		 0 	&\quad\quad	 		 \frac{1}{4}(\delta^{i}_{k}\delta^{j}_{l} 	 +\delta^{j}_{k}\delta^{i}_{l} )	&\quad\quad		0	&\quad\quad		 0 	&			
\\
	&\quad\quad		&\quad\quad		&\quad\quad	     &\quad\quad	          &\,	&\, 	&\quad\quad		
\\
0	&\quad\quad		 0 	&		-\frac{1}{4}(\delta^{i}_{k}\delta^{j}_{l} +\delta^{i}_{l}\delta^{j}_{k})	&\quad\quad			0	&\quad\quad		0	&\quad\quad		0	&\quad\quad	 	            
\\
	& \,	&\quad\quad			&\,        &\,          &\,	&\,	&\,
\\
0	&\quad\quad		0	&\quad\quad	 	0	&\quad\quad	 	0	&\quad\quad	 	0	&\quad\quad	 	0	&\quad\quad	 				
\\
0	&\quad\quad		0	&\quad\quad		0	&\quad\quad		0	&\quad\quad		0	&\quad\quad		0	&\quad\quad	  
\\
0	&\quad\quad	 	0	&\quad\quad	 	0	&\quad\quad		0	&\quad\quad		-\delta^{i}_{j}		&\quad\quad	 	0	
\\
0	&\quad\quad		0	&\quad\quad	 	-\frac{1}{2}(\delta^{k}_{i}\partial_{j}+\delta^{k}_{j}\partial_{i})		&\quad\quad	 	0	&\quad\quad		0	&\quad\quad		-\delta^{i}_{j}	
\end{array}
\right.\nonumber\\\nonumber\\
&& \left. 
\begin{array}{cccccccc}
0 & 0 \\
0& 0& \\
 0& \frac{1}{2} (\delta^k_i \partial_ j + \delta^k_ j \partial_i) \\
0 & 0 \\
\delta^i_ j & 0 \\
 0 &\delta_j^i\\
 0 & 0 \\
 0 & 0 \\
\end{array}
\right)\delta^{3}(x-y),
\end{eqnarray*}
where we can identify the following generalized FJ  brackets by means of 
\begin{equation}
\{\xi_{i}^{(1)}(x),\xi_{j}^{(1)}(y)\}_{FJ}\equiv\left(f_{ij}^{(1)}\right)^{-1}, 
\end{equation}
hence, the following FJ brackets arise 
\begin{eqnarray}
\{E_{ij}, P^{kl}\}_{FJ} &=& \frac{1}{4}\left(\delta^{k}_{i}\delta^{l}_{j} + \delta^{k}_{j}\delta^{l}_{i}\right)\delta^{3}(x-y), \\
\{B_{ij}, \Pi^{kl}\}_{FJ} &=& \frac{1}{4}\left(\delta^{k}_{i}\delta^{l}_{j} + \delta^{k}_{j}\delta^{l}_{i}\right)\delta^{3}(x-y),
\end{eqnarray}
and hence  Dirac's brackets and the  FJ ones coincide to each other. Furthermore, we have commented above that in the FJ framework there is not a classification between the constraints in first class and second class; in the FJ scheme  the counting of degrees of freedom is carry out as follows; there are 20 symplectic  variables given by $(E_{ij}, P^{ij}, B_{ij}, \Pi^{ij})$ and there are 6 constraints and 2 reducibility conditions, hence, there are 4  independent constraints and at the end one obtains eight physical degrees of freedom, such as it was obtained within the Dirac formalism. 
%%%%%%%%%%%%%%%%%%%%%%
\section{Symplectic analysis of standard linearized gravity}
Now, we will develop the symplectic analysis of the action (\ref{Llin}), and then we will compare the final results with those obtained in previous sections. We perform first the 3+1  decomposition  
\begin{gather}
\mathcal{L}=\dot{h}_{ij} \left[\frac{1}{4}\dot{h}_{kl} \left(\eta^{ki}\eta^{lj}-\eta^{ij}\eta^{kl} \right) +\frac{1}{2} \left(\partial^{j}h^{0i}+\partial^{i}h^{0j} \right)- \eta^{ij}\partial^{k}h^{0}_{\hphantom{0}k}\right] - \frac{1}{2}\partial_{i}h_{0j}\partial^{i}h^{0j}-  \frac{1}{4}\partial_{i}h_{jk}\partial^{i}h^{jk} \nonumber \\+\frac{1}{2}\partial_{i}h^{0}_{\hphantom{0}0}\partial^{i}h^{j}_{\hphantom{j}j}+\frac{1}{4}\partial_{i}h^{j}_{j}\partial^{i}h^{k}_{\hphantom{k}k}-\frac{1}{2}\partial_{i}h^{ij}\partial_{j}h^{0}_{\hphantom{0}0}   -\frac{1}{2}\partial_{i}h^{ij}\partial_{j}h^{k}_{\hphantom{k}k}+\frac{1}{2}\partial_{i}h_{j0}\partial^{j}h^{0i}+\frac{1}{2}\partial^{i}h_{jk}\partial^{j}h^{ik}. 
\label{lin2}
\end{gather}
By introducing the momenta given by 
\begin{gather}
\Pi^{mn}=\frac{1}{2}\dot{h}_{ij}\left( \eta^{mi}\eta^{nj}-\eta^{ij}\eta^{mn} \right) +\frac{1}{2} \left(\partial^{n}h^{0m}+\partial^{m}h^{0n} \right)- \eta^{mn}\partial^{k}h^{0}_{\hphantom{0}k}, 
\end{gather}
the Lagrangian acquires the following symplectic form 
\begin{gather}
\mathcal{L}^{(0)}=\dot{h}_{ij}\Pi^{ij}- \mathcal{V}^{(0)},  
\end{gather}
where $\mathcal{V}^{(0)}$ is the symplectic potential  
\begin{eqnarray}
\mathcal{V}^{(0)} &=& \Pi_{ij}\Pi^{ij}-\frac{1}{2}\eta_{kl}\eta_{ij}\Pi^{kl}\Pi^{ij}-2\partial_{j}h^{0}_{\hphantom{0}i}\Pi^{ij}- \frac{1}{4}\partial_{i}h_{jk}\partial^{i}h^{jk}-  
\frac{1}{2}\partial_{i}h^{0}_{\hphantom{0}0}\partial^{i}h^{j}_{\hphantom{j}j}- \frac{1}{4}\partial_{i}h^{j}_{\hphantom{j}j}\partial^{i}h^{k}_{\hphantom{k}k}\nonumber \\ &+& \frac{1}{2}\partial_{i}h^{ij}\partial_{j}h^{0}_{\hphantom{0}0} + \frac{1}{2}\partial_{i}h^{ij}\partial_{j}h^{k}_{\hphantom{k}k}- \frac{1}{2}\partial_{i}h_{jk}\partial^{j}h^{ik}.
\end{eqnarray}
On the other hand, from the symplectic Lagrangian (\ref{lin2}) it is possible to identify the following symplectic variables  and 1-forms respectively 
\begin{gather}
\xi^{(0)}=\left( h_{00},h_{0i},h_{ij},\Pi^{ij} \right), \\
a^{(0)}=\left(0,0,\Pi^{ij},0\right),
\end{gather}
thus, the symplectic matrix (\ref{37}) takes the form 
 \begin{eqnarray}
 f_{ij}^{(0)}(x,y)=  
 \left(
   \begin{matrix}
0 & 0 & 0  & 0   \\ \\
0 & 0 & 0 & 0    \\ \\ 
0 & 0 &  0 & -\frac{1}{2}\left(\delta^{l}_{i} \delta^{m}_{j} + \delta^{m}_{i} \delta^{l}_{j}  \right) \\ \\
0 & 0 & \frac{1}{2}\left(\delta^{i}_{l} \delta^{j}_{m} + \delta^{j}_{l} \delta^{i}_{m}  \right) & 0  \\ \\
   \end{matrix} 
\right)\delta^{3}(x-y),
   \end{eqnarray}
We observe that the symplectic matrix  is singular and this  means that there are constraints. It is important to note the difference between the alternative action analyzed in previous sections and the standard action. In fact, in the former the symplectic matrix was not singular, and the constraints where added by hand. In the later, there are constraints, which  will be different with  respect to  the nonstandard theory.    In order to identify the FJ constraints, we calculate the zero-modes of the symplectic matrix. These modes are given by 
\begin{gather*}
v_{1}^{(0)}=\left(v^{h_{00}},0,0,0,0,0 \right), \\
v_{2}^{(0)}=\left(0,v^{h_{0i}},0,0,0,0 \right), 
\end{gather*}
where  $v^{h_{00}}$, $v^{h_{0i}}$ are arbitrary functions.  Thus, the constraints will be obtained from  the contraction of the null vectors and the variation of the symplectic potential, this is,  $v^{i(0)}\frac{\delta \mathcal{V}^{(0)}\left(\xi^{(0)} \right)}{\delta \xi^{(0)i}}=0$,  \cite{F2}. In this manner, the following  FJ constraints arise 
\begin{gather}
\Omega^{(0)}_{1} = \int d^{3}x \ v_{1}^{(0)l} \ \frac{\delta}{\delta \xi^{(0)l}} \int d^{3}y \ V^{(0)} \left( \xi^{(0)} \right) = \int d^{3}x \ v^{h_{00}} \frac{\delta}{\delta h_{00}} \int d^{3}y V^{(0)} \Rightarrow \nonumber \\
\Omega^{(0)}_{1}= \frac{1}{2} \int d^{3}x \ v^{h_{00}} \left( \nabla^{2}h^{j}_{\hphantom{j}j}-\partial_{i}\partial_{j}h^{ij} \right)=0 \nonumber \\
 \Omega^{(0)}_{1}=  \frac{1}{2} \left( \nabla^{2}h^{j}_{\hphantom{j}j}-\partial_{i}\partial_{j}h^{ij} \right), \label{cons1}
\end{gather}
\begin{gather}
\Omega^{(0)}_{2} = \int d^{3}x \ v_{2}^{(0)l} \ \frac{\delta}{\delta \xi^{(0)l}} \int d^{3}y \ V^{(0)} \left( \xi^{(0)} \right) = \int d^{3}x \ v^{h_{0i}} \frac{\delta}{\delta h_{0i}} \int d^{3}y V^{(0)} \Rightarrow \nonumber  \\
\Omega^{(0)}_{2}=\int d^{3}x \ v^{h_{0i}} \partial_{j}\Pi^{ij}=0, \nonumber  \\
\Omega^{(0)i}_{2}=\partial_{j}\Pi^{ij}.  \label{cons1}
\end{gather}
Furthermore, in order to determine if there are more constraints, we demand consistency conditions, thus we  construct the following system \cite{F2}
\begin{gather}
\bar{f_{ij}}\xi^{(0)j}=Z_{i},
\end{gather}
where 
 $$
 \bar{f_{ij}}=  
 \left(
   \begin{matrix}
f_{ij}^{(0)} \\ \\
\frac{\delta \Omega_{1}^{0}}{\delta \xi^{(0)j}} \\ \\
\frac{\delta \Omega_{2}^{0}}{\delta \xi^{(0)j}}
   \end{matrix} 
\right),
   $$ with 
   $$ 
Z_{i}=
\left( 
\begin{matrix}
\frac{\delta V^{(0)}}{\delta \xi^{(0)i}} \\ \\
0 \\ \\
0
\end{matrix}
\right). 
   $$   
In this manner,  the matrix  $\bar{f_{ij}}$ takes the following form 
\begin{eqnarray}
\bar{f_{ij}}=
\left(
\begin{matrix}
0 & 0 & 0  & 0   \\ \\
0 & 0 & 0 & 0    \\ \\ 
0 & 0 &  0 & -\frac{1}{2}\left(\delta^{l}_{i} \delta^{m}_{j} + \delta^{m}_{i} \delta^{l}_{j}  \right) \\ \\
0 & 0 & \frac{1}{2}\left(\delta^{i}_{l} \delta^{j}_{m} + \delta^{j}_{l} \delta^{i}_{m}  \right) & 0  \\ \\
0 & 0 & \frac{1}{2} \left(\nabla^{2}\eta^{lm}-\partial^{l}\partial^{m} \right)  & 0 \\ \\
0 & 0 & 0 &  \frac{1}{2}\left(\partial_{m}\delta^{i}_{l}+\partial_{l}\delta^{i}_{m}  \right) \\ \\
\end{matrix}
\right)\delta^{3}(x-y), \label{35}
\end{eqnarray}
and 
$$
Z_{i}=
\left(
\begin{matrix}
\frac{1}{2} \left( \nabla^{2}h^{j}_{\hphantom{j}j}-\partial_{i}\partial_{j}h^{ij} \right) \\ \\
\partial_{j}\Pi^{ij}  \\ \\
\frac{1}{2} \left[ \eta^{ij} \nabla^2h^{0}_{\hphantom{0}0} - \partial^i \partial^j h^{0}_{\hphantom{0}0} \right] +  \frac{1}{2} \left[ \eta^{ij} \nabla^2h^{k}_{\hphantom{k}k} + \partial^i \partial^j h^{k}_{\hphantom{k}k} \right] - \frac{1}{2} \left[ \nabla^2h^{ij} + \eta^{ij} \partial^l \partial^k h_{lk} \right]  \\ \\
2\Pi_{lm}-\eta_{lm}\Pi^{j}_{\hphantom{j}j}-\partial_{m}h^{0}_{\hphantom{0}l}-\partial_{l}h^{0}_{\hphantom{0}m} \\ \\
0 \\ \\
0
\end{matrix}
\right)
$$
We can observe that the matrix (\ref{35}) is  singular, therefore there are zero-modes. The zero-modes of that matrix are given by
\begin{gather}
\bar{v_{1}}= \left( 0,0,0, -\left( \nabla^{2}\eta^{lm}-\partial^{l}\partial^{m} \right),\left(\delta_{l}^{i}\delta_{m}^{j}+\delta_{l}^{j}\delta_{m}^{i} \right),0 \right), \nonumber  \\
\bar{v_{2}}= \left( 0,0,\left( \partial_{m}\delta^{i}_{l}+\partial_{l}\delta^{i}_{m}  \right),0,0,\left(\delta_{i}^{l}\delta_{j}^{m}+\delta_{i}^{m}\delta_{j}^{l}\right) \right), \label{nul}
\end{gather}
thus, in order to determine if there are more FJ constraints, we calculate the contraction of the null vectors (\ref{nul}) with  $Z_{i}$ \cite{F2},  and   from that contraction  we can observe that there are no more FJ constraints because  the result is a combination of constraints  
\begin{gather}
\bar{v_{1}}^lz_{l}=2\partial_{l}\partial_{m}\Pi^{lm}=2\partial_{l}\Omega_{2}^{(0)l}=0,  \\
\bar{v_{2}}^iz_{i}=\partial^{i}\left( \nabla^{2}h^{j}_{\hphantom{j}j}-\partial_{i}\partial_{j}h^{ij} \right) =2\partial^{i}\Omega_{1}^{(0)}=0. 
\end{gather}
Furthermore,  we will  add the information of the FJ constraints to the action via Lagrange multipliers, namely $\alpha$ and   $\beta$, thus we construct a new symplectic Lagrangian   
\begin{gather}
\mathcal{L}^{(1)}=\dot{h}_{ij}\Pi^{ij} - \mathcal{V}^{(0)} \Big|_{\Omega_{1}^{(0)},\Omega_{2}^{(0)}=0}-\Omega_{1i}^{(0)}\dot{\alpha^{i}}-\Omega_{2i}^{(0)}\dot{\beta^{i}},  \nonumber \\
=\dot{h}_{ij}\Pi^{ij}-\frac{1}{2} \left( \nabla^{2}h^{j}_{\hphantom{j}j}-\partial_{i}\partial_{j}h^{ij} \right)\dot{\alpha}-\partial_{j}\Pi^{ij}\dot{\beta}_{i}-\Pi^{ij}\Pi_{ij}+\frac{1}{2}\eta_{ij}\eta_{kl}\Pi^{ij}\Pi^{kl}-  \nonumber \\ 
\frac{1}{4}\partial_{i}h_{jk}\partial^{i}h^{jk}+\frac{1}{4}\partial^{i}h^{j}_{\hphantom{j}j}\partial_{i}h^{k}_{\hphantom{k}k}-\frac{1}{2}\partial_{i}h^{ij}\partial_{j}h^{k}_{\hphantom{k}k}+\frac{1}{2}\partial_{i}h_{jk}\partial^{j}h^{ik}.\label{simp2}
\end{gather}
In this manner, from (\ref{simp2}) we identify the following new set of symplectic variables  
\begin{gather}
\xi^{(1)}= \left(h_{ij},\Pi^{ij},\alpha, \beta_i \right), \\
a^{(1)}=(\Pi^{ij},0,-\frac{1}{2} \left( \nabla^{2}h^{j}_{\hphantom{j}j}-\partial_{i}\partial_{j}h^{ij} \right),-\partial_{j}\Pi^{ij}).
\end{gather}
with these variables we calculate the new symplectic matrix 
\begin{eqnarray}
{\scriptsize
f_{ij}^{(1)}=
\left(
\begin{matrix}
 0  & -\frac{1}{2}\left(\delta_{i}^{l}\delta_{j}^{m}+\delta_{i}^{m}\delta_{j}^{l} \right) & \frac{1}{2} \left(\nabla^{2}\eta^{lm}-\partial^{l}\partial^{m} \right) & 0 \\ \\
 \frac{1}{2}\left(\delta_{l}^{i}\delta_{m}^{j}+\delta_{l}^{j}\delta_{l}^{i} \right)  & 0 & 0 & \frac{1}{2}\left(\partial_{m}\delta^{i}_{l}+\partial_{l}\delta^{i}_{m}  \right) \\ \\
  -\frac{1}{2} \left(\nabla^{2}\eta^{ij}-\partial^{i}\partial^{j} \right) & 0  & 0 & 0  \\ \\
  0  &- \frac{1}{2}\left(\partial_{i}\delta^{m}_{j}+\partial_{j}\delta^{m}_{i}  \right) & 0 & 0 \\ \\
\end{matrix} \label{ma2-}
\right)} \delta^3(x-y).
\end{eqnarray}
We can observe that the symplectic matrix (\ref{ma2-}) is singular, however, we have showed that there are no more constraints. Thus, this result indicate that linearized gravity  is a gauge theory,  as expected. In order to obtain a symplectic tensor, we need to fixing the gauge. We will use first  the temporal gauge, then we will work with the coulomb-like gauge, and  in both scenarios a symplectic tensor will be obtained. 
In this manner, by using the temporal gauge, we consider  that $h_{00}=0$ and $h_{0i}=0$, this implies that $\dot{\alpha}=0$,   $\dot{\beta}^{i}=0$,  and  we will  add the  Lagrange multipliers $\Sigma$, $\Gamma^i$,   enforcing this  gauge choice.  Thus, we obtain the following symplectic Lagrangian 
\begin{gather}
\mathcal{L}^{(2)}=\dot{h}_{ij}\Pi^{ij}+ \left[ \Sigma -\frac{1}{2} \left( \nabla^{2}h^{j}_{\hphantom{j}j}-\partial_{i}\partial_{j}h^{ij} \right) \right]\dot{\alpha}+ \left[\Gamma^i -\partial_{j}\Pi^{ij}\right]\dot{\beta}_{i}-\Pi^{ij}\Pi_{ij}+\frac{1}{2}\eta_{ij}\eta_{kl}\Pi^{ij}\Pi^{kl}-  \nonumber \\ 
\frac{1}{4}\partial_{i}h_{jk}\partial^{i}h^{jk}+\frac{1}{4}\partial^{i}h^{j}_{\hphantom{j}j}\partial_{i}h^{k}_{\hphantom{k}k}-\frac{1}{2}\partial_{i}h^{ij}\partial_{j}h^{k}_{\hphantom{k}k}+\frac{1}{2}\partial_{i}h_{jk}\partial^{j}h^{ik}, \label{eq43}
\end{gather}
where we identify the following symplectic variables and the following   1-forms respectively 
\begin{gather}
\xi^{(2)}= \left(h_{ij},\Pi^{ij},\alpha, \beta_i, \Sigma, \Gamma \right), \\
a^{(2)}=(\Pi^{ij},0,\left[ \Sigma -\frac{1}{2} \left( \nabla^{2}h^{j}_{\hphantom{j}j}-\partial_{i}\partial_{j}h^{ij} \right) \right], \Gamma^i-\partial_{j}\Pi^{ij}, 0, 0), 
\end{gather}
with these variables we calculate the corresponding  symplectic matrix, given by 
\begin{eqnarray}
{\scriptsize
f_{ij}^{(2)}=
\left(
\begin{matrix}
 0  & -\frac{1}{2}\left(\delta_{i}^{l}\delta_{j}^{m}+\delta_{i}^{m}\delta_{j}^{l} \right) & \frac{1}{2} \left(\nabla^{2}\eta^{lm}-\partial^{l}\partial^{m} \right) & 0& 0& 0 \\ \\
 \frac{1}{2}\left(\delta_{l}^{i}\delta_{m}^{j}+\delta_{l}^{j}\delta_{l}^{i} \right)  & 0 & 0 & \frac{1}{2}\left(\partial_{m}\delta^{i}_{l}+\partial_{l}\delta^{i}_{m}  \right) &0 &0 \\ \\
  -\frac{1}{2} \left(\nabla^{2}\eta^{ij}-\partial^{i}\partial^{j} \right) & 0  & 0 & 0 &-1 &0  \\ \\
  0  &- \frac{1}{2}\left(\partial_{i}\delta^{m}_{j}+\partial_{j}\delta^{m}_{i}  \right) & 0 & 0 & 0& - \delta^i_ j \\ \\
  0& 0& 1& 0& 0& 0 \\ \\
  0& 0& 0& \delta^i_ j& 0& 0
\end{matrix} \label{ma2-2}
\right)} \delta^3(x-y).
\end{eqnarray}
we  observe that  $f_{ij}^{(2)}$ is not singular, and therefore it is invertible. The inverse is given by 
\begin{eqnarray}
{\scriptsize
{f_{ij}^{(2)}}^{-1}=
\left(
\begin{matrix}
 0  & \frac{1}{2}\left(\delta_{i}^{l}\delta_{j}^{m}+\delta_{i}^{m}\delta_{j}^{l} \right) & 0 & 0& 0& \frac{1}{2} (\delta_j ^k \partial_i + \delta_i ^k \partial_j ) \\ \\
- \frac{1}{2}\left(\delta_{l}^{i}\delta_{m}^{j}+\delta_{l}^{j}\delta_{l}^{i} \right)  & 0 & 0 &0 & \frac{1}{2}\left( \eta^{ij} \nabla^2 - \partial^i\partial^j  \right) &0 \\ \\
  0 & 0  & 0 & 0 &1 &0  \\ \\
  0  &0 & 0 & 0 & 0&  \delta^i_ j \\ \\
  0& - \frac{1}{2}\left( \eta^{ij} \nabla^2 - \partial^i\partial^j  \right) & -1& 0& 0& 0 \\ \\
  -\frac{1}{2} (\delta_j ^k \partial_i + \delta_i ^k \partial_j ) & 0& 0&-  \delta^i_ j& 0& 0 
\end{matrix} \label{ma2-2-2}
\right)} \delta^3(x-y), 
\end{eqnarray}
thus, it is possible to identify the following generalized FJ brackets 
\begin{equation}
\{\xi_{i}^{(2)}(x),\xi_{j}^{(2)}(y)\}_{FJ}\equiv\left(f_{ij}^{(2)}\right)^{-1}, 
\end{equation}
where the relevant brackets are given by 
\begin{equation}
\{h_{ij} (x), \Pi^{kl}(y) \} =  \frac{1}{2}\left(\delta_{i}^{k}\delta_{j}^{l}+\delta_{i}^{l}\delta_{j}^{k} \right) \delta^{3} (x-y).
\end{equation}
Now, it is well-known  that in Dirac's terminology, to work with the temporal gauge implies to convert the primary first class constraints into second class ones. However, there are a remanent of   first class constraints. In this manner, if  we wish to convert all first class constraints into second class ones, we need to fix a different gauge; namely  the Coulomb gauge. In fact, in \cite{barcelos} it was performed the Dirac analysis  by using the Coulomb-like gauge, hence,   we will reproduce all these results by means a different approach. First, we will add to the Lagrangian (\ref{eq43}) the following Coulomb-like gauge  $\partial^j h_{ij}=0$, the  momentum gauge $ \Pi^i{_{i}}=0$, and the consequent Lagrange multipliers enforcing these gauge conditions, namely,  $\Lambda^i$, $\Upsilon$ 
\begin{gather}
\mathcal{L}^{(3)}=\dot{h}_{ij}\Pi^{ij}+ \left[ \Sigma -\frac{1}{2} \left( \nabla^{2}h^{j}_{\hphantom{j}j}-\partial_{i}\partial_{j}h^{ij} \right) \right]\dot{\alpha}+ \left[\Gamma^i -\partial_{j}\Pi^{ij}\right]\dot{\beta}_{i}-\Pi^{ij}\Pi_{ij}+\frac{1}{2}\eta_{ij}\eta_{kl}\Pi^{ij}\Pi^{kl}  \nonumber \\ 
- \Pi^i{_{i}} \Upsilon - \partial^j h_{ij} \Lambda ^i-\frac{1}{4}\partial_{i}h_{jk}\partial^{i}h^{jk}+\frac{1}{4}\partial^{i}h^{j}_{\hphantom{j}j}\partial_{i}h^{k}_{\hphantom{k}k}-\frac{1}{2}\partial_{i}h^{ij}\partial_{j}h^{k}_{\hphantom{k}k}+\frac{1}{2}\partial_{i}h_{jk}\partial^{j}h^{ik}, \label{eq50}
\end{gather}
from this symplectic Lagrangian, we identify the following symplectic variables and the following 1-forms respectively 
\begin{gather}
\xi^{(3)}= \left(h_{ij},\Pi^{ij},\alpha, \beta_i,  \Lambda ^i,  \Upsilon,  \Sigma, \Gamma^i \right), \\
a^{(3)}=(\Pi^{ij},0,\left[ \Sigma -\frac{1}{2} \left( \nabla^{2}h^{j}_{\hphantom{j}j}-\partial_{i}\partial_{j}h^{ij} \right) \right], \Gamma^i-\partial_{j}\Pi^{ij}, -\partial^j h_{ij}, - \Pi^i{_{i}},  0, 0), 
\end{gather}
thus, the symplectic  matrix has the form 
\begin{eqnarray}
{\scriptsize
f_{ij}^{(3)}=
\left(
\begin{matrix}
 0  & -\frac{1}{2}\left(\delta_{i}^{l}\delta_{j}^{m}+\delta_{i}^{m}\delta_{j}^{l} \right) &-  \frac{1}{2} \left(\nabla^{2}\eta^{lm}-\partial^{l}\partial^{m} \right) & 0& -\frac{1}{2} (\delta_i ^j\partial^l + \delta_i ^l\partial^j )& 0& 0 &0  \\ \\
 \frac{1}{2}\left(\delta_{l}^{i}\delta_{m}^{j}+\delta_{l}^{j}\delta_{l}^{i} \right)  & 0 & 0 &- \frac{1}{2}\left(\partial_{m}\delta^{i}_{l}+\partial_{l}\delta^{i}_{m}  \right) &0 &- \eta_{ij}& 0 & 0 \\ \\
  \frac{1}{2} \left(\nabla^{2}\eta^{ij}-\partial^{i}\partial^{j} \right) & 0  & 0 & 0 &0 &0 &-1& 0    \\ \\
  0  & \frac{1}{2}\left(\partial_{i} \delta^{m}_{j}+\partial_{j}\delta^{m}_{i}  \right) & 0 & 0 & 0&0 & 0&  -\delta^i_ j \\ \\
  \frac{1}{2} (\delta_i ^j \partial^l + \delta_i ^l\partial^j )& 0&0& 0& 0& 0&0 &0 \\ \\
0& \eta_{ij}& 0 & 0 & 0& 0&0&0 \\ \\
0& 0& -1 &0 & 0& 0&0&0 \\ \\
  0& 0& 0 &  \delta^i_ j & 0& 0&0&0 \\ \\
\end{matrix} \label{ma2-23}
\right)} \delta^3(x-y), \nonumber \\
\end{eqnarray}
we realize that the above matrix is not singular, and its inverse is given by
\begin{eqnarray}
{\scriptsize
{f_{ij}^{(3)}}^{-1}=
\left(
\begin{matrix}
 0  & A & (\eta_{ij}- \frac{\partial_i \partial_j}{ \nabla^2} )\frac{1}{\nabla^2} & 0& -B& 0& 0 &0  \\ \\
 -A  & 0 & 0 &-C &0 & \frac{1}{2}( \eta^{ij}- \frac{\partial^i \partial^j}{ \nabla^2}  ) & 0 & 0 \\ \\
-  (\eta_{ij}- \frac{\partial_i \partial_j}{ \nabla^2} )\frac{1}{\nabla^2} & 0  & 0 & \frac{\partial_i}{\nabla^4} &0 &- \frac{1}{\nabla^2} &1& 0    \\ \\
  0  & C&  -\frac{\partial_i}{\nabla^4}   & 0 & (\delta^i _j- \frac{\partial^i \partial_j}{ \nabla^2} )\frac{1}{\nabla^2} &0 & 0& \frac{\partial_i \partial^j}{2 \nabla^2}\\ \\
  B& 0&0& - (\delta^i _j- \frac{\partial^i \partial_j}{ \nabla^2} )\frac{1}{\nabla^2}  & 0& 0&0 &0 \\ \\
0&  - \frac{1}{2}(\eta^{ij}- \frac{\partial^i \partial^j}{ \nabla^2} ) & \frac{1}{\nabla^2} & 0 & 0& 0&0&0 \\ \\
0& 0& -1&0 & 0& 0&0&0 \\ \\
  0& 0& 0 &-\frac{\partial_i \partial^j}{2 \nabla^2} & 0& 0&0&0 \\ \\
\end{matrix} \label{ma2-23}
\right)} \delta^3(x-y), \nonumber \\
\end{eqnarray}
where we have defined 
\begin{eqnarray}
A&=& \frac{1}{2}\left(\delta_{i}^{l}\delta_{j}^{k}+\delta_{i}^{k}\delta_{j}^{l} \right)- \left[ \delta_i ^l  \partial_j \partial^k+  \delta_i ^k \partial_j \partial^l + \delta_j ^k  \partial_i \partial^l +  \delta_j^l  \partial_i \partial^k \right]\frac{1}{2\nabla^2} -\frac{1}{2} \eta^{kl}\eta_{ij} + \frac{1}{2} (\eta_{ij} \partial^k \partial^l + \eta^{kl} \partial_i \partial_j )\frac{1}{\nabla^2}  \nonumber \\
&+ &\frac{\partial_i \partial_j \partial^k \partial^l}{2 \nabla^4}, \nonumber\\
B&=&  \frac{\partial^k \partial_i \partial_j}{\nabla^4} - \left[ (\delta_{i}^{k} \partial_j+\delta_{j}^{k} \partial_i \right]\frac{1}{\nabla^2} ,  \nonumber \\
C&=& \left[ \eta^{kl} \partial_i + \frac{\partial_i \partial^k \partial^l}{\nabla^2}  \right] \frac{1}{2\nabla^2} - \left[ \delta_i ^l \partial^k + \delta^k_i \partial^l \right] \frac{1}{2\nabla^2}. 
\end{eqnarray}
In this manner, we can identify the nontrivial FJ brackets by means of 
\begin{eqnarray}
\{\xi_{i}^{(3)}(x),\xi_{j}^{(3)}(y)\}_{FJ}\equiv\left(f_{ij}^{(3)}\right)^{-1},
\end{eqnarray}
thus, we find 
\begin{eqnarray}
\{ h_{ij} (x), \Pi^{kl}(y)\} &=& \Big[ \frac{1}{2}\left(\delta_{i}^{l}\delta_{j}^{k}+\delta_{i}^{k}\delta_{j}^{l} \right)-\frac{1}{2} \left[ \delta_i ^l  \partial_j \partial^k+  \delta_i ^k \partial_j \partial^l + \delta_j ^k  \partial_i \partial^l +  \delta_j^l  \partial_i \partial^k \right]\frac{1}{\nabla^2} -\frac{1}{2} \eta^{kl}\eta_{ij} \nonumber \\ 
 &+& \frac{1}{2} (\eta_{ij} \partial^k \partial^l + (\eta_{kl} \partial_i \partial_j )\frac{1}{\nabla^2}  +  \frac{\partial_i \partial_j \partial^k \partial^l}{2 \nabla^4}\Big] \delta^3(x-y), \nonumber\\
\end{eqnarray}
these brackets reproduce exactly those obtained  in \cite{barcelos} by using the Dirac method, and we have reproduced the same  brackets by means of a different way.\\ 
On the other hand, we can observe the differences between the nonstandard theory and the standard one; the standard linearized theory is a gauge theory, while the nonstandard theory is not. Furthermore, for both theories   the generalized brackets between the fields are different, in addition, we would like to comment that our results are absent in the literature and   they are an   extension of those reported in \cite{barcelos}.  
\section{Conclussions}
In this paper,  the canonical and symplectic analysis  for an alternative action describing gravity were performed.  With respect to  Dirac's formalism, we found the constraints of the theory, which turned out to be  of second class, then the Dirac brackets were constructed. In addition, we found  that the Hamiltonian  is of first class and therefore it correspond to an observable; these facts  were  not reported in \cite{r1, r2}. Furthermore, the results were  reproduced using the FJ formalism: We constructed a symplectic tensor, then the  generalized FJ brackets were found, we showed that Dirac's and FJ brackets coincide to each other. In this manner,   we have confirmed  by an alternative way that  the  action is not a gauge theory and   its physical degrees of freedom are eight. In order to complete our analysis,  we added  the symplectic analysis of standard linearized gravity. \\
We finish this paper with some points to remark. If we define a system in terms of its  equations of motion, then an  infinity number of Hamiltonian structures can be  defined for the same system \cite{Ge}. This fact present a problem for  singular systems  because one  could propose  several actions yielding the same equations of motion, but the symmetries between these  different actions could be different just  as it was showed  in the present  analysis. Thus, in order to study any new proposal for a physical system,   it is mandatory to put attention in  the symmetries  beyond its  equations of motion \cite{al}. 
\newline
\newline
\newline
\noindent \textbf{Acknowledgements}\\[1ex]
This work was supported by CONACyT under Grant No.CB-$2014$-$01/ 240781$. We would like to  thank R. Cartas-Fuentevilla for discussion on the subject and reading of the manuscript.

\end{document}